\documentclass[10pt,aps,prc,twocolumn,showpacs,showkeys,amsmath,floatfix,superscriptaddress]
{revtex4-1}
\usepackage{color,graphicx}
\usepackage{CJK}
\usepackage{mathptmx}                
\usepackage{dcolumn}                 
\usepackage{bm}                      
\usepackage{ulem,soul}

\begin{document}
\begin{CJK*}{GBK}{song}

\title{Reinvestigation of the excited states in the proton emitter $^{151}$Lu: particle-hole excitations across the $N=Z=64$ subshell}
\author{F.~Wang}
\affiliation{School of Physics and Nuclear Energy Engineering,
Beihang University, Beijing 100191, China}

\author{B.H.~Sun}
\email{Corresponding author: bhsun@buaa.edu.cn}
\affiliation{School of Physics and Nuclear Energy Engineering,
Beihang University, Beijing 100191, China}

\author{Z.~Liu}
\email{liuzhong@impcas.ac.cn}
\affiliation{Institute of Modern Physics, Chinese Academy of Sciences, Lanzhou 730000, China}
\affiliation{Department of Physics, University of Surrey, Guildford, Surrey GU2 7XH, UK}%

\author{C.~Qi}
\email{chongq@kth.se}
\affiliation{KTH, Alba Nova University Center, SE-10691 Stockholm, Sweden}

\author{L.H.~Zhu}
\affiliation{School of Physics and Nuclear Energy Engineering, Beihang University, Beijing 100191, China}

\author{C. Scholey}
\affiliation{University of Jyvaskyla, Department of Physics, P.O. Box 35, FI-40014 University of Jyvaskyla, Finland}

\author{S.F. Ashley}
 \affiliation{Department of Physics, University of Surrey, Guildford, Surrey GU2 7XH, UK}%

\author{L. Bianco}
 \affiliation{Department of Physics, Oliver Lodge Laboratory, University of Liverpool, Liverpool L69 7ZE, UK}%


\author{I.J. Cullen}
\affiliation{Department of Physics, University of Surrey, Guildford, Surrey GU2 7XH, UK}

\author{I.G. Darby}
\affiliation{Department of Nuclear Sciences and Applications, International Atomic Energy Agency, A-1400 Vienna, Austria}

\author{S. Eeckhaudt}
\affiliation{University of Jyvaskyla, Department of Physics, P.O. Box 35, FI-40014 University of Jyvaskyla, Finland}

\author{A.B. Garnsworthy}
\affiliation{TRIUMF, 4004 Wesbrook Mall, Vancouver, British Columbia V6T 2A3, Canada}

\author{W. Gelletly}
\affiliation{Department of Physics, University of Surrey, Guildford, Surrey GU2 7XH, UK}

\author{M.B. Gomez-Hornillos}
\affiliation{Universitat Polit\'{e}cnica de Catalunya (UPC), 08034 Barcelona, Spain}

\author{T. Grahn}
\affiliation{University of Jyvaskyla, Department of Physics, P.O. Box 35, FI-40014 University of Jyvaskyla, Finland}

\author{P.T. Greenlees}
\affiliation{University of Jyvaskyla, Department of Physics, P.O. Box 35, FI-40014 University of Jyvaskyla, Finland}

\author{D.G. Jenkins}
 \affiliation{Department of Physics, University of York, Heslington, York, UK YO10 5DD, UK}%

\author{G.A. Jones}
\affiliation{Department of Physics, University of Surrey, Guildford, Surrey GU2 7XH, UK}

\author{P. Jones}
\affiliation{iThemba LABS, National ResearchFoundation, PO Box 722, Somerset West, South Africa}

\author{D.T. Joss}
\affiliation{Department of Physics, Oliver Lodge Laboratory, University of Liverpool, Liverpool L69 7ZE, UK}

\author{R. Julin}
\affiliation{University of Jyvaskyla, Department of Physics, P.O. Box 35, FI-40014 University of Jyvaskyla, Finland}

\author{S. Juutinen}
\affiliation{University of Jyvaskyla, Department of Physics, P.O. Box 35, FI-40014 University of Jyvaskyla, Finland}

\author{S. Ketelhut}
\affiliation{University of Jyvaskyla, Department of Physics, P.O. Box 35, FI-40014 University of Jyvaskyla, Finland}

\author{S. Khan}
\affiliation{Schuster Building, School of Physics and Astronomy, University of Manchester, Manchester M13 9PL, UK}

\author{A. Kishada}
\affiliation{Schuster Building, School of Physics and Astronomy, University of Manchester, Manchester M13 9PL, UK}

\author{M. Leino}
\affiliation{University of Jyvaskyla, Department of Physics, P.O. Box 35, FI-40014 University of Jyvaskyla, Finland}

\author{M. Niikura}
\affiliation{CNS, University of Tokyo, Tokyo 351-0100, Japan}

\author{M. Nyman}
\affiliation{European Commission, Joint Research Centre, Institute for Reference Materials and Measurements, Retieseweg 111, B-2440 Geel, Belgium}

\author{R.D. Page}
\affiliation{Department of Physics, Oliver Lodge Laboratory, University of Liverpool, Liverpool L69 7ZE, UK}

\author{J. Pakarinen}
\affiliation{Department of Physics, Oliver Lodge Laboratory, University of Liverpool, Liverpool L69 7ZE, UK}

\author{S. Pietri}
\affiliation{GSI Helmholtzzentrum f\"{u}r Schwerionenforschung, D-64291 Darmstadt, Germany}

\author{Zs. Podoly\'{a}k}
\affiliation{Department of Physics, University of Surrey, Guildford, Surrey GU2 7XH, UK}

\author{P. Rahkila}
\affiliation{University of Jyvaskyla, Department of Physics, P.O. Box 35, FI-40014 University of Jyvaskyla, Finland}

\author{S. Rigby}
\affiliation{Department of Physics, Oliver Lodge Laboratory, University of Liverpool, Liverpool L69 7ZE, UK}

\author{J. Sar\'{e}n}
\affiliation{University of Jyvaskyla, Department of Physics, P.O. Box 35, FI-40014 University of Jyvaskyla, Finland}

\author{T. Shizuma}
\affiliation{Japan Atomic Energy Agency, Tokai, Ibaraki 319-1195, Japan}

\author{J. Sorri}
\affiliation{University of Jyvaskyla, Department of Physics, P.O. Box 35, FI-40014 University of Jyvaskyla, Finland}

\author{S. Steer}
\affiliation{Department of Physics, University of Surrey, Guildford, Surrey GU2 7XH, UK}

\author{J. Thomson}
\affiliation{Department of Physics, Oliver Lodge Laboratory, University of Liverpool, Liverpool L69 7ZE, UK}

\author{N.J. Thompson}
\affiliation{Department of Physics, University of Surrey, Guildford, Surrey GU2 7XH, UK}

\author{J. Uusitalo}
\affiliation{University of Jyvaskyla, Department of Physics, P.O. Box 35, FI-40014 University of Jyvaskyla, Finland}

\author{P.M. Walker}
\affiliation{Department of Physics, University of Surrey, Guildford, Surrey GU2 7XH, UK}

\author{S. Williams}
\affiliation{Department of Physics, University of Surrey, Guildford, Surrey GU2 7XH, UK}

\begin{abstract}
The excited states of the proton emitter $^{151}$Lu were reinvestigated in a recoil-decay tagging experiment at the Accelerator Laboratory of the University of  Jyv\"{a}skyl\"{a} (JYFL). The level scheme built on the ground state of $^{151}$Lu was updated with five new $\gamma$-ray transitions.
Large-scale shell model calculations
were carried out in the model space consisting of the neutron and proton orbitals $0g_{7/2}$, $1d_{5/2}$, $1d_{3/2}$, $2s_{1/2}$, and $0h_{11/2}$ with the optimized monopole interaction in order to interpret the experimental level scheme of $^{151}$Lu.
It is found that the excitation energies of states above the $27/2^-$ and $23/2^+$ isomeric levels in $^{151}$Lu can be sensitive to excitations from $g_{7/2}$ and $d_{5/2}$ to single-particle orbitals above $N=Z=64$.
\end{abstract}

\date{\today}

\maketitle

\section{Introduction}
Investigations of proton-emitting nuclei can provide invaluable information on nuclear structure beyond the proton drip line~\cite{p1997,blank2008nuclear,pfutzner2012radioactive}, such as masses and single particle orbitals. The experimental studies of proton emitters, however, are extremely difficult due to very low production cross sections as well as the presence of very strong contamination from other reaction products. This is witnessed by the existence of a limited number of $\gamma$-spectroscopy studies on proton-emitting nuclei.

$^{151}$Lu is the first case known to have ground state (g.s.) proton decay. It was observed in the fusion reaction $^{96}$Ru($^{58}$Ni, 1p2n)$^{151}$Lu at the
velocity filter at SHIP/GSI~\cite{zpa1982},
with the proton energy and half-life measured to be 1233 keV and 85(10) ms, respectively.
It was interpreted as $h_{11/2}$ g.s. proton decay.
Later, a 1310(10) keV proton decay with a much shorter half-life of 16(1) $\mu$s was found in $^{151}$Lu and assigned as proton decay from the $d_{3/2}$ isomer, the experimental spectroscopic factor for which was found to be much reduced~\cite{prc1999}.
More recent work~\cite{wang2017,prc15}, with the refined proton-decay data for the $d_{3/2}$ isomer in $^{151}$Lu, has resolved the discrepancy in spectroscopic factors and the extracted proton formation factor indicates no
significant hindrance for this isomeric proton decay.

The excited states of $^{151}$Lu were studied in different laboratories using the very selective recoil decay tagging (RDT) technique~\cite{prc1998,lz,plb13,prc15}. The level scheme of $^{151}$Lu was much extended in the recent experiment at JYFL~\cite{plb13,prc15}. The lifetimes of a few excited states including 15/2$^{-}$, the first excited \mbox{state} feeding the proton-emitting g.s., were measured using the recoil-distance Doppler-shift method~\cite{plb13,prc15}. The comparison between the measured lifetimes and theoretical calculations using the non-adiabatic strong-coupling model suggested a mild oblate deformation for the g.s. of $^{151}$Lu~\cite{plb13}.
A tentative level scheme built on the proton decaying $d_{3/2}$ \mbox{isomer} was proposed in Ref.~\cite{prc15}, but not all of the transitions could be confirmed in our recent work~\cite{wang2017}.

In this article, we report the updated level scheme of $^{151}$Lu on top of the 11/2$^-$ ground state from the same RDT experiment performed at JYFL as in Ref.~\cite{wang2017}.
The results are interpreted in term of large-scale shell model calculations.
The paper is organized as follows.
The experimental setup and the results are presented in Section II.
In Section III, the results are discussed by examining the systematics of the high-spin states in $N=80$ isotones, and also by comparison with large-scale shell model calculations.
A summary is given in Section IV.

\section{Experiment and Results}
\begin{figure}
\includegraphics[width=0.45\textwidth]{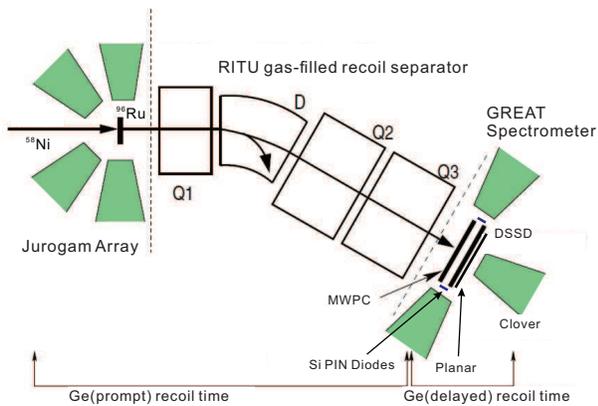}\\
\caption{(Color online) Schematic view of the recoil-decay tagging setup in the experiment.}\label{fig:fig1}
\end{figure}

A schematic view of the experimental setup is shown in Fig.~\ref{fig:fig1}. It consists of the JUROGAM array~\cite{juro} at the target position, the gas-filled recoil separator RITU~\cite{ritu,ritu1} and the GREAT spectrometer at the focal plane of RITU.

The excited states of $^{151}$Lu were populated with the $^{96}$Ru($^{58}$Ni, $1p2n$)$^{151}$Lu fusion-evaporation reaction in an experiment performed at JYFL. Part of the experimental results was reported in Ref.~\cite{wang2017}.
Prompt $\gamma$ rays emitted in the fusion-evaporation reactions were detected by the JUROGAM array comprising 43 HPGe detectors.
After a time of flight of about 0.6 $\mu$s in the gas-filled recoil separator RITU, the evaporation residues were implanted into a pair of 300-$\mu$m thick double-sided silicon-strip detectors (DSSSDs), which can record signals of recoils implanted and the energies of protons, $\alpha$ particles, $\beta$ rays and conversion electrons that were emitted.
The triggerless data acquisition system, total data readout (TDR), was used in our experiment.
In the TDR, each channel was running independently and the registered signal was time-stamped with a global 100 MHz clock.
This allows 
one to correlate the prompt $\gamma$ ray with implantation and \mbox{subsequent} decays (proton decays in the present work) within a given pixel of DSSSDs. The data were analyzed with GRAIN~\cite{grain}. More details of the experimental setup and data acquisition system can be found in Refs.~\cite{ritu,great,tdr,wang2017}.

\begin{figure}
  \includegraphics[width=8.5cm]{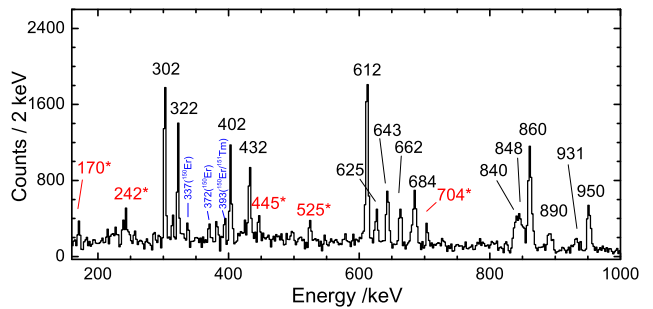}\\
\caption{\label{fig:fig2}(Color online) Background-subtracted total projection of $\gamma-\gamma$ matrix tagged within 250 ms with proton decays from the ground state of $^{151}$Lu. The $\gamma$-ray peaks indicated with an asterisk are $\gamma$ transitions observed for the first time in this work. $\gamma$-rays from the main contaminants $^{150}$Er and $^{151}$Tm are also indicated in the spectrum.
}
\end{figure}

A total yield of $2\times 10^5$ for the $^{151}$Lu g.s. protons was recorded.
These protons were then used to tag prompt $\gamma$ rays emitted at the target position.
A $\gamma-\gamma$ matrix was constructed from the $\gamma$ rays correlated with the g.s. protons decaying within 250 ms ($\approx3T_{1/2}$) after implantation. The total projection of this $\gamma-\gamma$ matrix is shown Fig.~\ref{fig:fig2}.

\begin{figure*}
  \includegraphics[width=0.65\textwidth]{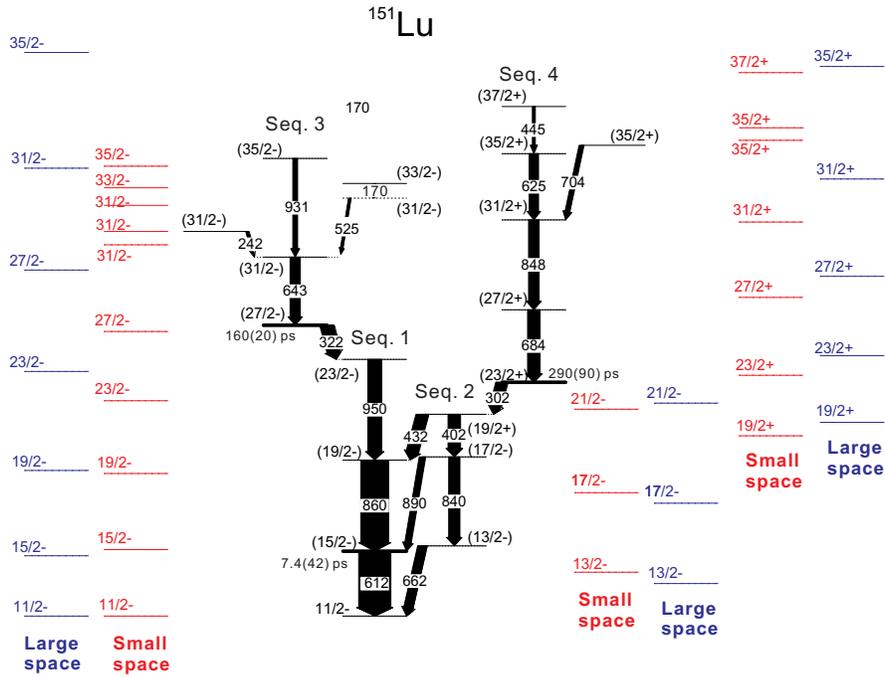}\\
\caption{\label{fig:fig3} (Color online) Level scheme based on the g.s. of $^{151}$Lu established in this work. Tentative spin-parity assignments are indicted in brackets. Theoretical calculations are done in the large-scale shell model. The half-lives for the ($15/2^-$), ($27/2^-$) and ($23/2^+$) states are from Ref.~\cite{plb13}.
}
\end{figure*}

All the $\gamma$ rays in coincidence with the g.s. proton decay of $^{151}$Lu observed previously ~\cite{plb13,prc15} are present in Fig.~\ref{fig:fig2}.
In addition, new $\gamma$-ray transitions (labelled with asterisks) are observed at energies of 170, 242, 445, 525 and 704 keV.
The energies, relative intensities of the $\gamma$-ray transitions identified are summarized in Table I, together with the tentative spin and parity assignments in brackets for the levels. 
The new level scheme of $^{151}$Lu, shown in Fig.~\ref{fig:fig3}, is proposed based on the $\gamma-\gamma$ coincidence relationships, energy sums, relative intensities and intensity balance.
Due to the low statistics it is not possible to assign the multipolarity from angular distributions.

\begin{table}
\begin{center}
\caption{ \label{tab1}  Energies and relative intensities for $\gamma$ transitions assigned to $^{151}$Lu. The relative intensity of the 612-keV transitions, feeding the g.s., is normalized to 100\%.}
\begin{tabular}{ccccccc}
\hline
$E_{\gamma}(keV)$  &  &$J^{\pi}_i$  & &  $J^{\pi}_f$&    &$I_{\gamma}(\%)$    \\
\hline
170.4(15)& &(33/2-)& &(31/2-)&  &10(3) \\
242.3(10)& &(31/2-)& &(31/2-)&  &8(4)\\
301.8(3)& &(23/2+)& &(19/2+)&  &44(2)  \\
322.3(4)& &(27/2-)& &(23/2-)&  &35(2) \\
401.8(6)& &(19/2+)& &(17/2-)&  &39(2) \\
431.7(6)& &(19/2+)& &(19/2-)&  &30(2) \\
445.2(12)& &(37/2+)& &(35/2+)&  &9(5)\\
524.6(10)& &(31/2-)& &(31/2-)&  &10(3) \\
612.3(4)& &(15/2-)& &11/2-&  &100(2) \\
625.3(5)& &(35/2+)& &(31/2+)&  &29(2) \\
642.6(5)& &(31/2-)& &(27/2-)&  &32(2) \\
662.1(6)& &(13/2-)& &11/2-&  &29(2) \\
684.4(5)& &(27/2+)& &(23/2+)&  &39(3) \\
703.8(10)& &(35/2+)& &(31/2+)&  &15(3) \\
840.1(12)& &(17/2-)& &(13/2-)&  &35(4) \\
847.5(12)& &(31/2+)& &(27/2+)&  &33(4)\\
860.3(5)& &(19/2-)& &(15/2-)&  &87(3) \\
890.1(10)& &(17/2-)& &(15/2-)&  &20(3) \\
930.9(10)& &(35/2-)& &(31/2-)&  &14(5) \\
950.3(6)& &(23/2-)& &(19/2-)&  &44(3) \\
\hline
\end{tabular}
\end{center}
\end{table}

The part of the level scheme built on the g.s. of $^{151}$Lu established in Ref.~\cite{plb13} is confirmed in the present work. The low-lying level pattern in $^{151}$Lu is expected to be similar to that of the neighboring odd-$Z$ $N=80$ isotones $^{145}$Tb~\cite{145tb} and $^{147}$Ho~\cite{Roth2001}, where the most strongly populated yrast levels are $15/2^-$,  $19/2^-$, and $23/2^-$ in order of increasing excitation \mbox{energy}. The level sequence formed by the 612-, 860- and 950-keV $\gamma$-ray transitions is assigned tentatively as the corresponding cascade in $^{151}$Lu.

\begin{figure}
  \includegraphics[width=0.5\textwidth]{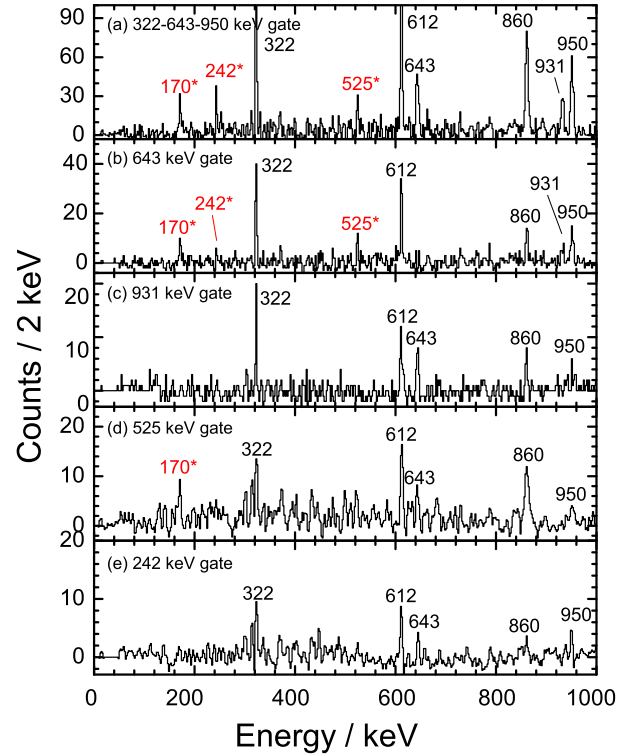}\\
\caption{\label{fig:fig4}(Color online) Background-subtracted prompt $\gamma-\gamma$ spectra: (a) 322-643-950 keV gated sum spectrum, (b) 643 keV gated spectrum, (c) 931 keV gated spectrum, (d) 525 keV gated spectrum, (e) 242 keV gated spectrum. The $\gamma$-ray peaks indicated with an asterisk are $\gamma$ transitions observed for the first time in this work.
}
\end{figure}

The transition sequence of 612, 860, 950, 322, 643 and 931 keV has been reported in Refs.~\cite{plb13,prc15} and can be clearly seen in the sum spectrum gated on 950, 322 and 643 keV $\gamma$ rays as shown in Fig.~\ref{fig:fig4}(a).
In addition, three weak peaks at 170, 242, 525 keV are present.
These three $\gamma$ rays can be seen in the spectrum gated on the 643-keV transition (see Fig.~\ref{fig:fig4}(b)), but are not visible in the spectrum gated on the 931-keV $\gamma$ ray (see Fig.~\ref{fig:fig4}(c)).
This suggests that they are in parallel with the 931-keV transition.
The 170- and 525-keV transitions are found in coincidence with each other, but not with the 242-keV $\gamma$-ray (see Fig.~\ref{fig:fig4}(d)(e)).

\begin{figure}
  \includegraphics[width=0.48\textwidth]{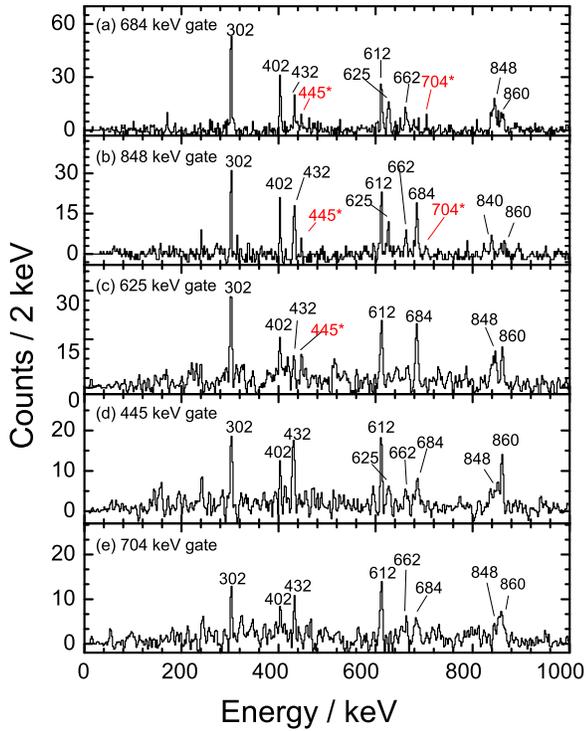}\\
\caption{\label{fig:fig5}(Color online) Same as Fig.~\ref{fig:fig4} but for (a) 684 keV gated spectrum, (b) 848 keV gated spectrum, (c) 625 keV gated spectrum, (d) 445 keV gated spectrum, (e) 704 keV gated spectrum. The $\gamma$-ray peaks indicated with an asterisk are $\gamma$ transitions observed for the first time in this work.
}
\end{figure}

\begin{figure}
  \includegraphics[width=8.5cm]{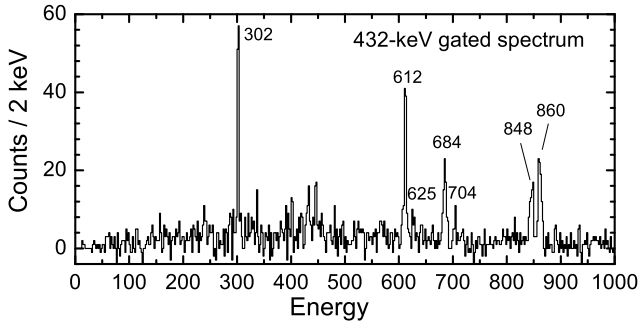}\\
\caption{\label{fig:fig6}Background-subtracted 432 keV gated spectrum, in which the 840-keV can be isolated from the 848-keV since 432-keV is not in coincidence with that of 840-keV transition.
}
\end{figure}

As shown in Fig.~\ref{fig:fig5}(a)(b), two new $\gamma$ rays at 445 and 704 keV are observed in the spectra gated on the 684- and 848-keV transitions of Seq. 4 in the level scheme.
The 704 $\gamma$ rays can be seen in the 848-keV gated spectrum but not the 625-keV gated spectrum (see Fig.~\ref{fig:fig5}(c)).
In addition, the 625-keV transition is not present in the 704-keV gated spectrum (see Fig.~\ref{fig:fig5}(e)), indicating that these two transitions are in parallel.
All the $\gamma$ rays except 704 keV of Seq. 4 can be seen in the spectrum gated on 445 keV (see Fig.~\ref{fig:fig5}(d)).
The 445-keV transition is placed on the top of Seq. 4 according to the relative intensity compared with the 625-keV transition in the spectrum gated on the 684-keV transition.
By gating the 432-keV transition, the 848-keV transition can be well resolved from that of 840-keV since 432-keV is not in coincidence with that of 840-keV transition, shown in Fig.\mbox{~\ref{fig:fig6}}. The 684-keV intensity, 39(3)\%, is slightly larger than that of 848-keV, 33(4)\%, in one standard deviation. The order of the 848- and 684-keV transitions is proposed on the basis of their relative intensities from the 432 keV gated spectrum, which is in agreement with Ref.\mbox{~\cite{plb13}} but opposite to that in Ref.\mbox{~\cite{prc15}}.

\section{Discussion}\label{sect:shell}

\subsection{Systematics of the high-spin states in $N=80$ isotones}
\begin{figure}
  \includegraphics[width=0.5\textwidth]{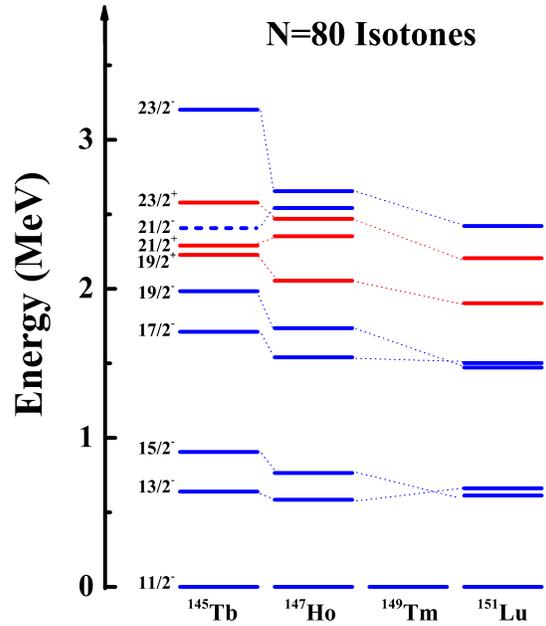}\\
\caption{\label{fig:fig7} (Color online) Level schemes of $^{151}$Lu and its lighter $N=80$, odd-$Z$ isotones. Data for $^{145}$Tb and $^{147}$Ho are taken from Refs.~\cite{145tb,Roth2001}, respectively. The dashed line implies the level with a tentative spin-parity assignment.
  }
\end{figure}

The systematics of the yrast states of the three odd-$Z$, $N=80$ isotones $^{145}$Tb, $^{147}$Ho and $^{151}$Lu are shown in Fig.~\ref{fig:fig7}.
No experimental information on $^{149}$Tm is available yet.
It is noted that the excitation energies of the yrast levels of $15/2^-$, $19/2^-$ and $23/2^-$ decrease with proton number $Z$.
The energy gaps between the $15/2^-$ and $13/2^-$, $19/2^-$ and $17/2^-$, and $23/2^-$ and $21/2^-$ levels decrease with increasing $Z$, implying that the $21/2^-$ level could be very close to 
the $23/2^-$ level in $^{151}$Lu.
By taking this into consideration, the level fed by the 302-keV transition and depopulated by the 402- and 432-keV transitions, may not be the $21/2^-$ state as suggested in Ref.~\cite{prc15}.
Also it can be seen that the $19/2^+$ and $23/2^+$ levels have a decreasing trend with the proton number.
The level fed by the 302-keV transition is now assigned as ($19/2^+$) while the 290-ps isomeric state as ($23/2^+$).

\subsection{Shell model interpretation}

The nuclear shell model is a fundamental approach to \mbox{study} the microscopic structure of, in principle, any nuclei of any shape. But in practice, due to the limit of computing capacity, \mbox{shell} model is successful only in medium and heavy nuclei with or close to spherical shape. With $N=80$ and small deformation, $^{151}$Lu is now within the reach of large-scale shell model study.
To understand the structure of the observed excited states of $^{151}$Lu, large-scale shell model configuration interaction calculations have been carried out in the model \mbox{space} consisting of the neutron and proton orbitals $0g_{7/2}$, $1d_{5/2}$, $1d_{3/2}$, $2s_{1/2}$, and $0h_{11/2}$ (denoted as $gdsh$ hereafter). That is, $^{151}$Lu is described as the coupling of 2 neutron holes and 11 proton holes (or 21 valence protons and 30 valence neutrons in the particle-particle channel). The \mbox{calculations} are done with a realistic nucleon-nucleon interaction with the monopole-interaction channel optimized as described in \mbox{Ref.}~\cite{Qi2012PRC}.
That interaction has been shown to reproduce very well the spectroscopic properties of Sn and heavier Sb, Te, I isotopes close to $Z=50$. However, it has not been tested in nuclei heavier than $Z=64$, which can be quite sensitive to the non-diagonal neutron-proton interaction matrix elements involving protons and neutron holes
across the $Z=N=64$ subshell closures.
Therefore, two sets of calculations are presented for the excited states as shown in Fig.~\ref{fig:fig3}. In the first case, the calculations were carried out in the full $gdsh$ model space by considering all possible particle excitations and the results are denoted as ``large-space'' in Fig.~\ref{fig:fig3}.
Calculations in a relatively smaller model space were also performed, by restricting the maximal number of particle-hole excitations
across the $N=Z=64$ subshell closures (i.e., neutron/proton particle excitations from the $d_{5/2}$ and $g_{7/2}$ orbitals to $s_{1/2}$, $d_{3/2}$ and $h_{11/2}$ or vice verse in terms of hole excitations) to two. This is feasible due to the fact that the low-lying states are dominated by the coupling of valence neutron holes and proton holes in the $d_{3/2}$, $s_{1/2}$ and $h_{11/2}$ shells, which are close to each other. The restriction also makes it easier to evaluate the nonyrast states. We have calculated the lowest three states for each spin/parity upto $J$=37. Part of the results are shown in Fig.~\ref{fig:fig3} and are labeled as ``small-space''.
However, even in this case the dimension of the bases is still quite large, at the order of $10^7$.
The results between these two calculations are close to each other for the lowest-lying states, and are in good agreement with the experimental data.
However, noticeable deviations start from 27/2$^-$ and 31/2$^+$ states, which indicates
that the large-space calculation seems to overestimate the particle/hole excitations across the $N=Z=64$ subshell for those states.
For example, as for the first $27/2^-$ state, the average numbers of protons that excited to above the $Z=64$ subshell closure are calculated to be 0.491 and 1.856 in large- and small-space calculations, respectively. The calculations indicate that those states are sensitive to the particle-hole excitations from the $d_{5/2}$ and $g_{7/2}$ orbitals to $s_{1/2}$, $d_{3/2}$ and $h_{11/2}$ orbitals, and can serve as a good test ground for the crossing subshell interactions mentioned above.

To understand the difference between the two sets of calculations, the occupancies of single-particle orbitals were calculated for all yrast states. The average number of the protons in the orbital $h_{11/2}$ is calculated to be around five.
It is noticed that for the large-space calculation, the average number of particles (or holes) in each orbital remains roughly the same for all states shown in Fig.~\ref{fig:fig3}. In relation to this, the yrast states in the large-space calculation show a rather collective structure with large in-band $E2$ transitions, indicating that those calculated states have similar intrinsic structure. The calculated spectroscopic quadrupole moments also remain practically the same for all yrast states. However, as indicated in the figure, the large-space calculation may have overestimated the energies of the higher lying states from $27/2^-$ and the crossing-subshell excitations. The crossing subshell proton-neutron interactions need to be adjusted in order to have a correct description of those higher-lying states. On the other hand, the small-space calculation, in which a weaker particle-hole excitation 
crossing the $N$=$Z$=64 subshell closures is explicitly imposed, seems to reproduce the experimental data better.

The present large-scale shell calculations are done within the so-called M-scheme where the total magnetic quantum number is conserved in the bases. For calculations in the smaller space, another advantage is that it is possible to project the wave function as a coupling of the proton group and neutron group with good angular momenta in the form $|\phi^p_{\pi}(J_{\pi})\otimes\phi^n_v(J_{v})\rangle$, where $J_{\pi}$ and $J_{v}$ denote the angular momenta of the protons and neutrons (see, e.g., Ref.~\cite{Qi2011PRC}), respectively. As expected, the $^{151}$Lu g.s. is dominated by the one quasi-particle configuration $|\phi^p_{\pi}(J_{\pi}=11/2^-)\otimes \phi^n_{v}(J_{v}=0^+)\rangle$. The wave functions for other low-lying states show a similar structure and are dominated by proton excitations. The next most important components correspond to the coupling of $J_{v}=2^+$ neutron-hole pair and proton states.

\begin{table*}
\begin{center}
\caption{ \label{tab2}The reduced transition probabilities for the 302-, 322- and 612-keV transitions in $^{151}$Lu under different multipolarity assumptions, the  small-space shell model results for $E2$ transitions are also listed.}
\begin{tabular}{ccccccccc}
\hline
$E_{\gamma}/keV$  &lifetime(ps)&$B(M1)\downarrow$($\mu^2_N$)  & $B(M1)\downarrow$(W.u.) &$B(E2)\downarrow$($e^2fm^4$)& $B(E2)\downarrow$(W.u.) & $B(E2)_{\rm theo}$($e^2fm^4$)&$B(E1)\downarrow$($e^2fm^2$) &$B(E1)\downarrow$(W.u.)    \\
\hline
302&290(90)&$7.1(22)\times10^{-3}$&$4.0(12)\times10^{-3}$&1125(349)&23(7)&265&$7.9(24)\times10^{-5}$&$4.3(13)\times10^{-5}$ \\
322&160(20)&$1.0(1)\times10^{-2}$&$6.0(7)\times10^{-3}$&1480(184)&31(34)&1712&$1.1(1)\times10^{-4}$&$6.4(8)\times10^{-5}$ \\
612&7.4(42)&$3.3(19)\times10^{-2}$&$1.9(10)\times10^{-2}$&1290(732)&27(15)&1508&$3.7(21)\times10^{-4}$&$2.0(11)\times10^{-4}$ \\
\hline
\end{tabular}
\end{center}
\end{table*}

\subsection{Negative Parity Band}
The lifetimes of the $15/2^-$ level and two high-lying states were measured in Ref.~\cite{plb13}. Assuming pure stretched transitions with no mixing from higher-order multipolarities, the reduced transition probabilities can be deduced~\cite{igs,Li2016} for the 612-, 322- and 302-keV transitions and the results are listed in Table~\ref{tab2} for different multipolarities.
The values are given in Weisskopf units (W.u.) as well.
The $B(E2)$ value for the 15/2$^-\rightarrow$11/2$^-$ 612-keV transition calculated from the small-space shell-model (1508 $e^2$fm$^4$) reproduces well the experimental one.

It is interesting to note that the pattern of the low-lying negative parity yrast states up to $23/2^-$ in $^{151}$Lu is similar to that of $^{155}$Lu \cite{155lu,2016lu155} with two neutrons above the $N=82$ shell closure.
Both are quite different from that of $^{153}$Lu~\cite{153lu} with frozen neutron degree of freedom ($N=82$). The three quasi-particle 27/2$^-$ yrast states are expected to lie slightly higher than 23/2$^-$ and to be isomeric in both $^{151}$Lu and $^{153}$Lu. The 23/2$^-$ and 27/2$^-$ states are calculated to be nearly degenerate in $^{153}$Lu, in agreement with the experimental data. On the other hand, the $25/2^-$ yrast state has been observed to be lower than the $27/2^-$ and even $23/2^-$ states in the nucleus $^{155}$Lu in relation to the enhanced anti-aligned neutron-proton interaction~\cite{2016lu155}.

The 27/2$^-$ level may be of particular interest as 27/2 is the maximum spin that can be formed from three protons in the $h_{11/2}$ orbital. Indeed, this fully aligned configuration is the leading component for the isomeric 27/2$^-$ state in $^{151}$Lu, bringing it down in excitation energy.

The levels in Seq. 3 are assigned as ($27/2^-$), ($31/2^-$) and ($35/2^-$).
Their excitation energies are well reproduced by the small-space shell-model calculations.
As expected the neutron hole pairs occupy mainly the $d_{3/2}$ and $s_{1/2}$ orbitals. The $B(E2)$ value for the 322-keV transition is extracted to be 1480 $e^2$fm$^4$, in good agreement with the value of 1712 $e^2$fm$^4$ predicted by the small space shell model.

The level depopulated by the newly observed 242-keV transition is assigned as ($31/2^-$) by comparing with the non-yrast results of small-space calculation.
The intensities of the 525- and 170-keV transitions in Table~\ref{tab1} have similar intensities within statistical uncertainties, whereas the order of the two transitions as well as the assignments are mainly determined by referring to the small-space calculation.

\subsection{Positive Parity Band}
The $B(E2)$ value extracted for the 302-keV transition is 1125 $e^2$fm$^4$, while it is 265 $e^2$fm$^4$ in the small-space shell-model calculation.
The levels above the 290-ps isomeric \mbox{state} agree very well with the small-space calculation, thus the spin and parity of the states in the positive-parity band are assigned by mainly referring to the theoretical small-space calculation.
In Refs.~\cite{plb13,prc15}, these levels were assigned as (25/2$^+$), (29/2$^+$), (33/2$^+$) by comparing with the nonadiabatic quasiparticle calculation. Our recent measurements ~\cite{wang2017} suggest that the proton-decay properties of $^{151}$Lu can be well explained without introducing deformation. It is thus expected that the low-lying spectrum of the nucleus can be well explained within the shell model framework without introducing cross-shell quadrupole-quadrupole correlations.
The levels depopulated by the 445-keV and 704-keV transitions are tentatively assigned as ($37/2^+$) and ($35/2^+$), respectively.
The shell-model calculation predicts strong E2 transitions between the 35/2$^+_{1,2}$ and 31/2$^+_1$ state with B(E2) around 860 $e^2$fm$^4$ for both cases. The E2 transitions from 37/2$^+_1$ to 35/2$^+_{1,2}$ states are calculated to be much weaker than above transitions, for which the calculated  $B(E2)$ are 88 and 194 $e^2$fm$^4$, respectively.

The positive-parity states are dominated by the excitation of one proton and one neutron-hole to $s_{1/2}$ and $d_{3/2}$ orbitals. The $E2$ transitions within the yrast bands are calculated to be pretty strong (1000-2000 $e^2$fm$^4$) except the transition 23/2$^+\rightarrow$19/2$^+$, which is five times weaker than 27/2$^+\rightarrow$23/2$^+$.
The leading component for the $19/2^+$ level corresponds to $|\phi^p_{\pi}(J_{\pi}=19/2^+)\rangle$, particularly the configuration $\pi(d^3_{3/2}h^4_{11/2})$, coupled to the $J=0$ neutron hole pair. However, the wave functions for the other positive-parity states given by the two calculations are quite different: The orbital $vh_{11/2}$ remains inactive in the large-space calculation, whereas the neutron-hole configurations $s^{-1}_{1/2}\otimes h^{-1}_{11/2}$ and $d^{-1}_{3/2}\otimes h^{-1}_{11/2}$ play important roles in the small-space calculation. As a result, the $B(E2)$ values for the 23/2$^+\rightarrow$19/2$^+$ transition predicted by the two calculations are significantly different: the $B(E2)$ value predicted by the small-space calculation is much smaller than the 4706 $e^2$fm$^4$ predicted by the large-space calculation.

\section{Summary}
In the present work, the excited states of the proton emitter $^{151}$Lu were reinvestigated in a RDT experiment.
The level scheme built on the g.s. of $^{151}$Lu has been extended and compared with large-scale shell model calculations
in the full gdsh model space (``large-space'') and a truncated model space with limited particle/hole excitations across the presumed $N=Z=64$ subshell closures (``small-space'').
It is found that the low-lying states including the isomeric $27/2^-$ state are dominated by proton excitations with some contribution from the coupling of proton excitations to the $J_{v}=2^+$ neutron-hole pair. The excitation energies of levels above the isomeric $27/2^-$ and $23/2^+$ states are found to
fit well with the small-space calculation. This indicated that the particle/hole excitations across the $N=Z=64$ subshell closures tend to be overestimated in the large-space calculation for states above the isomeric $23/2^-$ and $31/2^+$ states. Further experimental results in the mass region are called for to constrain the crossing subshell neutron-proton interaction in the shell model calculations.

\begin{acknowledgments}
This work has been supported by the National Natural Science Foundation of China under
Nos. 11475014, 11435014, 11405224, 11205208, 11675225, 11635003 and U1632144,
the National Key Research and Development program(MOST 2016YFA0400501),
the "100 Talented Project" of the Chinese Academy of Sciences,
the EU 6th Framework programme "Integrating
Infrastructure Initiative - Transnational
Access", Contract Number: 506065 (EURONS),  the Academy of
Finland under the Finnish Centre of Excellence Programme 2006-2011
(Nuclear and Accelerator Based Physics Programme at JYFL), and the United Kingdom Science and Technology Facilities Council.
CS(209430) and PTG(111965) acknowledge the support of the Academy of Finland.
CQ is supported by the Swedish Research Council (VR) under grant Nos. 621-2012-3805, and
621-2013-4323 and the G\"oran Gustafsson foundation. The computations were performed on resources
provided by the Swedish National Infrastructure for Computing (SNIC) at PDC, KTH, Stockholm.
We would also like to thank Bao-An Bian, Lang Liu, and Zheng-Hua Zhang for valuable discussions.
\end{acknowledgments}

\bibliography{mybib}

\end{CJK*}
\end{document}